\documentclass[runningheads,11pt,oribibl]{llncs}
\usepackage{graphicx}
\usepackage{amsmath}
\usepackage{amssymb}
\usepackage{path}
\usepackage{xspace}
\usepackage[font=small, format=hang, labelfont=bf]{caption}
\usepackage[subrefformat=parens, labelfont=default]{subfig}
\usepackage{hyperref}
\usepackage{times}
\usepackage{wrapfig}
\usepackage{fullpage}

\title{Progress on Partial Edge Drawings}

\authorrunning{Bruckdorfer et al.}

\author{Till~Bruckdorfer \inst1 \and Sabine Cornelsen \inst2 \and
  Carsten~Gutwenger \inst3 \and Michael~Kaufmann \inst1 \and
  Fabrizio~Montecchiani \inst4 \and Martin~N\"ollenburg \inst5 \and
  Alexander~Wolff \inst6}

\institute{Universit\"at T\"ubingen, Germany \and Universit\"at
  Konstanz, Germany \and Universit\"at Dortmund, Germany \and
  Universit\`a degli Studi di Perugia, Italy \and Institut f\"ur Theoretische
  Informatik, KIT, Germany \and Lehrstuhl f\"ur Informatik~I,
  Universit\"at W\"urzburg, Germany}

\newcommand{\remove}[1]{}
\newenvironment{pf}{\begin{proof}}{\qed\end{proof}}

\newcommand{\eps}{\ensuremath{\varepsilon}\xspace}
\newcommand{\Oin}{\ensuremath{O_\mathrm{in}}\xspace}
\newcommand{\Oout}{\ensuremath{O_\mathrm{out}}\xspace}
\newcommand{\kopt}{\ensuremath{k_\mathrm{opt}}\xspace}
\newcommand{\Sleft}{\ensuremath{S_\mathrm{left}}\xspace}

\graphicspath{{figures/}}

\def\withcomments{
   \newcounter{mycommentcounter}
   \def\comment##1{\refstepcounter{mycommentcounter}%
    \ifhmode%
     \unskip%
     {\dimen1=\baselineskip \divide\dimen1 by 2 %
       \raise\dimen1\llap{\tiny -\themycommentcounter-}}\fi%
     \marginpar{\renewcommand{\baselinestretch}{0.8}%
       \footnotesize [\themycommentcounter]: \raggedright ##1}}
   }
\withcomments

\begin{document}

\maketitle

\begin{abstract}
  Recently, a new way of avoiding crossings in straight-line drawings
  of non-planar graphs has been investigated.  The idea of
  \emph{partial edge drawings} (PED) is to drop the middle part of
  edges and rely on the remaining edge parts called \emph{stubs}.  We
  focus on a symmetric model (SPED) that requires the two stubs of an
  edge to be of equal length.  In this way, the stub at the other
  endpoint of an edge assures the viewer of the edge's existence.  We
  also consider an additional homogeneity constraint that forces the
  stub lengths to be a given fraction~$\delta$ of the edge lengths
  ($\delta$-SHPED).  Given length and direction of a stub, this model
  helps to infer the position of the opposite stub.

  We show that, for a fixed stub--edge length ratio~$\delta$, not all
  graphs have a $\delta$-SHPED.  Specifically, we show that $K_{241}$
  does not have a $1/4$-SHPED, while bandwidth-$k$ graphs always have
  a $\Theta(1/\sqrt{k})$-SHPED.  We also give bounds for complete
  bipartite graphs.  Further, we consider the problem \textsc{MaxSPED} where
  the task is to compute the SPED of maximum total stub length that a
  given straight-line drawing contains.  We 
  present an efficient solution for 2-planar drawings and a
  2-approximation algorithm for the dual problem.
\end{abstract}

\section{Introduction}

In the layout of graphs, diagrams, or maps, one of the central
problems is to avoid the interference of elements such as crossing
edges in graph drawings or overlapping labels on maps.  This is a form
of \emph{visual clutter}.  Clutter avoidance is the objective of a
large body of work in graph drawing, information visualization, and
cartography.  In this paper, we treat a specific aspect of clutter
avoidance; we focus on \emph{completely} removing edge crossings in
straight-line drawings of non-planar graphs.  Clearly, this is not
possible in any of the traditional graph drawing styles that insist on
connecting the geometric representations of two adjacent vertices
(e.g., small disks) 
by a closed Jordan curve (e.g., segments of straight lines).
In such drawings of non-planar graphs, some pairs of edge
representations must cross (or overlap).
This is a serious problem when displaying dense graphs.

\paragraph{Previous Work.}

Becker et al.\ \cite{bew-vnd-TVCG95} have taken a rather radical
approach to escape from this dilemma.  They wanted to visualize
network overload between the 110 switches of the AT\&T long distance
telephone network in the U.S.\ on October~17, 1989, when the San
Francisco Bay area was hit by an earthquake.  They used straight-line
segments to connect pairs of switches struck by overload; the width of
the segments indicated the severeness of the overload.  Due to the
sheer number of edges of a certain width, the underlying map of the
U.S.\ was barely visible.  They solved this problem by drawing only a
certain fraction (roughly 10\%) of each edge; the part(s) incident to
the switch(es) experiencing the overload.  We call these parts the
\emph{stubs} of an edge.  The resulting picture is much clearer; it
shows a distinct east--west trend among the edges with overload.

Peng et al.\ \cite{plcp-srrpg-PV12} used splines to bundle edges,
e.g., in the dense graph of all U.S.\ airline connections.  In
order to reduce clutter, they increase the transparency of edges
towards the middle.  They compared their method to other edge bundling
techniques \cite{hw-fdebg-CGF09,ghms-maebv-PV11}, concluding that
their method, by emphasizing the stubs, is better in revealing
directional trends.

Burch et al.\ \cite{bvkw-epdld-GD11} recently investigated the usefulness
of partial edge drawings of \emph{directed} graphs.  They used a single
stub at the source vertex of each edge.  They did a user 
study (with 42 subjects) which showed that, for one of the three tasks
they investigated (identifying the vertex with highest out-degree),
shorter stubs resulted in shorter completion times \emph{and} smaller
error rates.  For the two other tasks (deciding whether a highlighted
pair of vertices is connected by a path of length one/two) the error
rate went up with decreasing stub length; 
there was just a small
dip in the completion time for a stub--edge length ratio of 75\%.

A similar, but less radical approach, is the use of edge
\emph{casing}.  Eppstein et al.\ \cite{ekms-estb-CGTA09} have
investigated how to optimize several criteria that encode the
above--below behavior of edges in given graph drawings.  They
introduce three models (i.e., legal above--below patterns) and
several objective functions such as minimizing the total number of
above--below switches or the maximum number of switches per edge.  For
some combinations of models and objectives, they give efficient
algorithms, for one they show NP-hardness; others are still open.
Edge casings were re-invented by Rusu et al.\ \cite{rfjr-ugpca-IV11}
with reference to Gestalt principles. 

Dickerson et al.~\cite{degm-cdvnp-JGAA05} proposed confluent drawings
to avoid edge crossings. In their approach, edges are drawn as locally
monotone curves; edges may overlap but not cross.

We build on and extend the work of Bruckdorfer and Kaufmann
\cite{bk-mecbe-FUN12} who formalized the problem of partial edge
drawings (PEDs) and suggested several variants. 
A PED is a straight-line drawing of a graph in which each edge is
divided into three segments: a middle part that is not drawn and the
two segments incident to the vertices, called stubs, that remain in the
drawing.  In this paper, we require all PEDs to be crossing-free,
i.e., that no two stubs intersect.  We consider stubs relatively open
sets, i.e., to not contain their endpoints.
In the symmetric case
(SPED), both stubs of an edge must be the same length; in the
homogeneous case (HPED), the ratio of stub length over edge length is
the same over all edges.  A $\delta$-SHPED is a symmetric homogeneous
PED with ratio~$\delta$.

\begin{figure}[tb]
  \subfloat[with crossings\label{sfg:crossings}]%
  {\quad\includegraphics[scale=1.4]{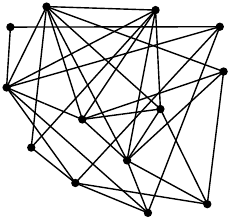}\quad}
  \hfill
  \subfloat[as a maxSPED\label{sfg:maxsped}]%
  {\quad\includegraphics[scale=1.4]{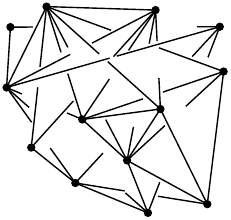}\quad}
  \hfill
  \subfloat[as~a~maxSHPED\label{sfg:maxshped}]%
  {\quad\includegraphics[scale=1.4]{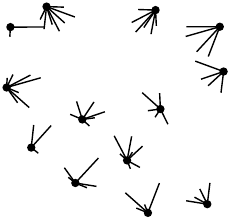}\quad}
  \caption{Various drawings of a 13-vertex graph.}
  \label{fig:example}
\end{figure}

Bruckdorfer and Kaufmann showed, among others, that $K_n$ (and thus,
any $n$-vertex graph) has a $1/\sqrt{4n/\pi}$-SHPED.  
They also proved that the $j$-th power of any subgraph of a triangular
tiling is a $1/(2j)$-SHPED.  
They introduced the optimization problem \textsc{MaxSPED} where the
aim is to maximize the total stub length (or \emph{ink}) in order to
turn a given geometric graph into a SPED.  They presented an integer
linear program for \textsc{MaxSPED} and conjectured that the problem
is NP-hard.  Indeed, there is a simple reduction from \textsc{Planar3SAT} 
\cite{ks-pc-12}.  Figure~\ref{sfg:maxsped} depicts a maxSPED of
the straight-line drawing in \figurename~\ref{sfg:crossings}, i.e.,
a SPED that is a solution to \textsc{MaxSPED}. 
We have slightly shrunken the stubs in the maxSPED
so that they do not touch. 
For comparison, \figurename~\ref{sfg:maxshped} depicts a SHPED with
maximum ratio~$\delta$.

\paragraph{Contribution.}

In this paper, we focus on the symmetric case. A pair of stubs of equal length
pointing towards each other at the opposite endpoints of an edge is, for
the viewer of a SPED, a valuable witness that the connection actually
exists.
If the drawing is additionally homogeneous, finding the other endpoint of a stub is made easier since its approximate distance can be estimated from the stub length.
\begin{itemize}
\item We show that not all graphs admit a 1/4-SHPED; see
  Sect.~\ref{sec:upper}.  Indeed, $K_n$ does not have a
  1/4-SHPED for any $n > 240$.  If we restrict
  vertices to be mapped to points in convex and one-sided convex
  position, the bound drops to~22 and~16, respectively.  Our
  proof technique carries over to other values of the stub--edge
  length ratio~$\delta$.
\item Recall that Bruckdorfer and
  Kaufmann~\cite{bk-mecbe-FUN12} showed that 
  $K_n$ (and thus, any $n$-vertex graph) has a
  $1/\sqrt{4n/\pi}$-SHPED.  We improve their result for specific graph
  classes, namely for complete bipartite graphs and for bandwidth-$k$
  graphs; see Sect.~\ref{sec:specific-classes}.  The latter we show
  to have $\Theta(1/\sqrt{k})$-SHPEDs independently of their sizes.
\item Then we turn to the optimization problem \textsc{MaxSPED};
  see Sect.~\ref{sec:speds}.  For the class of 2-planar graphs, we
  can solve the problem efficiently; given a 2-planar drawing of a
  2-planar graph with $n$
  vertices, our algorithm runs in $O(n\log n)$ time.  For general
  graphs, we have a 2-approximation algorithm with respect to the dual
  problem \textsc{MinSPED}: minimize the amount of ink that has to be
  erased in order to turn a given drawing into a SPED.
\end{itemize}

\paragraph{Notation.}

In this paper, we always identify the vertices of the given graph with
the points in the plane to which we map the vertices.
The graphs we consider are undirected; we use $uv$ as shorthand for
the edge connecting~$u$ and~$v$.  
If we refer to the stub~$uv$ then we mean the piece of the edge~$uv$
incident to~$u$; the stub~$vu$ is incident to~$v$.

\section{Upper Bounds for Complete Graphs}
\label{sec:upper}

In this section, we show that not any graph can be drawn as a $1/4$-SHPED.
Note that $1/4$ is an interesting value since it balances the drawn and the
erased parts of each edge. Yet, our proof techniques generalize to
$\delta$-SHPEDs for arbitrary but fixed $0 < \delta < 1/2$. We start with a simple proof
for the scenario where we insist that vertices are mapped to specific point
sets, namely point sets in convex or one-sided convex position. We say that a
convex point set is \emph{one-sided} if its convex hull contains an edge of a
rectangle enclosing the point set.


\begin{theorem}
  \label{thm:K_17}
  There is no set of 17 points in one-sided convex position
  on which the graph $K_{17}$ 
  can be embedded as $1/4$-SHPEDs, respectively.
\end{theorem}

\begin{pf}
  We assume, to the contrary of
  the above statement, that there is a set $P$ of 17 points in
  one-sided convex position that admits an embedding of $K_{17}$ as a
  $1/4$-SHPED. 
  Consider the edge $e=uv$ that witnesses the one-sidedness of~$P$.
  We can choose our coordinate system such that~$u = (0,0)$, $v=(1,0)$
  and all other points lie above~$e$.
  We split the area above~$e$ into twelve interior-disjoint
  vertical strips of equal width, see \figurename~\ref{fig:k17}.

  We first show that the union 
  of the six innermost strips contains at most six points of~$P$. 
  Otherwise there would be a strip~$S$ that contains two
  points~$a$ and~$b$ of $P$.
  Let~$a$ be the one closer to~$u$.
  Since~$S$ is one of the six innermost strips,
  the stub~$av$ intersects the right boundary of~$S$ (below the stub~$bv$),
  and the stub~$bu$ intersects the left boundary of~$S$ (below the
  stub~$au$).  
  Point~$a$ lies above stub $bu$ and point~$b$ lies above stub~$av$.  
  Hence, stubs~$av$ and~$bu$ intersect.
	
  \begin{figure}[tb]
    \centering
    \includegraphics{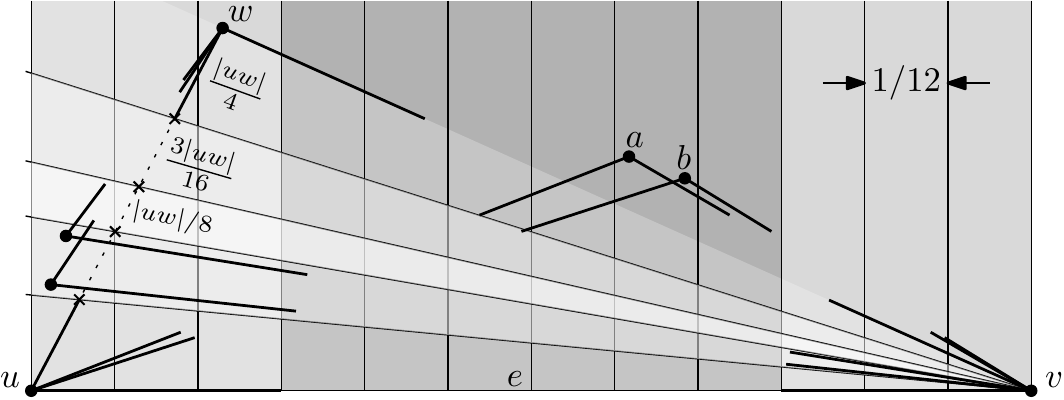}
    \caption{Sketch of the argument why no 17 points in one-sided
      convex position can be used to embed $K_{17}$ as a $1/4$-SHPED.} 
    \label{fig:k17}
  \end{figure}

  So at least eleven points of $P$ must lie in the union of the three
  leftmost and the three rightmost strips. 
  We may assume that the union~\Sleft of the three leftmost
  strips contains at least six points. 
  Let $w$ be the rightmost point in $P \cap \Sleft$.
  We subdivide the edge~$uw$ into five pieces whose lengths are $1/4$,
  $3/16$, $1/8$, $3/16$, and $1/4$ of the length of~$uw$.  Each piece
  contains its endpoint that is closer to one of the endpoints
  of~$uw$.  The innermost piece contains both of its endpoints.  Now
  consider the cones with apex~$v$ spanned by the five pieces of~$uw$.
  We claim that no cone contains more than one point.

  Our main tool is the following.  Let~$t$ be a point in $(P \cap
  \Sleft) \setminus \{u,w\}$.  Then the stub~$tv$ intersects the right
  boundary of~\Sleft and, hence, also the edge~$uw$ that separates $P
  \cap \Sleft$ from $P \setminus \Sleft$.
  It remains to note that in each cone, any point has a stub to~$u$
  or~$w$ (whichever is further away from the cone) that intersects the
  boundary of the cone.
  %
\end{pf}

%
Theorem~\ref{thm:K_17} can be used to derive a first upper bound on
general point sets as follows. 

\begin{corollary}
  \label{thm:generalpoints}
  For any $n > \binom{30}{15}$, the graph~$K_n$ does not admit a
  $1/4$-SHPED.
\end{corollary}

\begin{pf}
  By a result of Erd{\H o}s and Szekeres~\cite{es-cpg-35}, any set of
  more than $\binom{2k - 4}{k-2}$ points in general position contains
  a subset of~$k$ points that form a one-sided convex set.  Combining
  this with Theorem~\ref{thm:K_17} and plugging in~$k=17$ yields the
  claimed bound.
\end{pf}

We now vastly improve upon the bound of Corollary~\ref{thm:generalpoints}.  
%
Let $P$ be the point set in the plane, and let~$l$ and~$r$ be the two
points on the convex hull that define the diameter 
of~$P$, which is
the largest distance between any two points.  We rotate~$P$ such that
the line~$lr$ is horizontal and $l$ is on the left-hand side.  
Now let~$R$ be the smallest enclosing axis-aligned rectangle that
contains~$P$, and let~$t$ and~$b$ be the top- and bottommost points
in~$R$, respectively.  Accordingly, let~$R_t$ be the part of~$R$ above
(and including)~$lr$ and let $R_b = R \setminus R_t$.
We consider the two rectangles separately and assume that the interior
of~$R_t$ is not empty.  (In our proof we argue, for any interior
point, using only its stubs towards the three boundary points~$l$,
$r$, and~$t$.)

\begin{figure}
  \centering
  \includegraphics{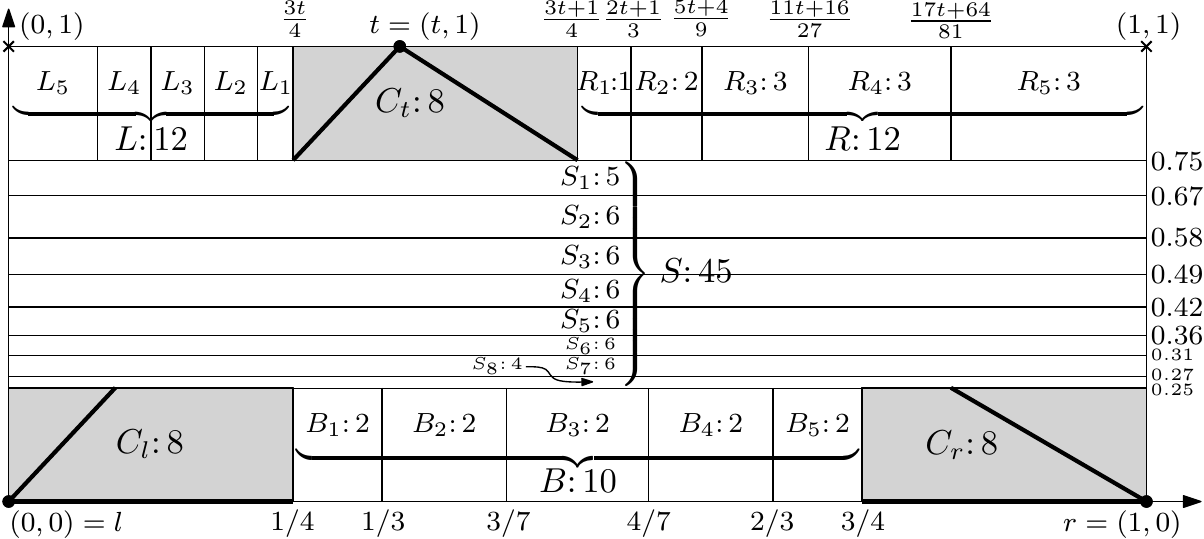}
  \caption{Partition of the enclosing rectangle $R_t$ into cells.  We
    have labeled each cell or group of cells with the maximum number
    of points that it can contain.}
  \label{fig:partition}
\end{figure}

We subdivide~$R_t$ into 26 cells
such that for each point in a cell the three stubs to $l$,
$r$, and~$t$ intersect the boundary of that cell; 
see \figurename~\ref{fig:partition}.
For each cell, we prove, in the remainder of this section, an upper bound 
on the maximum number of points it can contain.  Summing up these
numbers (see again \figurename~\ref{fig:partition}), we get a bound of~121
points in total.  Since we may have a symmetric subproblem below~$lr$, 
we double this number, subtract~2 because of
double-counting~$l$ and~$r$, and finally get the following theorem. 

\begin{theorem}
  \label{thm:improved}
  For any $n > 240$, the graph $K_n$ does not admit a $1/4$-SHPED.
\end{theorem}

We now prove Theorem~\ref{thm:improved} by upperbounding, for each
cell in \figurename~\ref{fig:partition}, the number of points it
contains.

For ease of presentation, we 
stretch $R_t$ in y-direction to make it a square. 
Clearly, this operation does not change the crossing properties.
We assume that the side length of~$R_t$ is~$1$.  We further
assume that the coordinates of~$l$, $r$, and~$t$, are
$(0,0)$, $(0,1)$, and $(t,1)$, respectively.
Note that, by the choice of~$r$ and~$l$, 
there are no other points on the left and right boundary of~$R_t$
(otherwise~$lr$ would not be the diameter of~$P$). 
By symmetry, we may further assume that $0 < t \leq 1/2$.
For a point~$p$, we call stub~$pt$ the \emph{upper stub} of~$p$, $pr$
its \emph{right stub}, $pl$ its \emph{left stub}, and both~$pr$
and~$pl$ its \emph{lower stubs}.  

For $p \in \{l,r,t\}$, let~$C_p \subset R_t$ be the axis-parallel
rectangle spanned by~$p$ and the endpoints of the two stubs that go
from~$p$ to the two other boundary points.  Note that~$C_l, C_r$, and~$C_t$
(all shaded in \figurename~\ref{fig:partition}) are squares of size
$1/4 \times 1/4$.


\paragraph{The middle strip.}

We first consider the middle strip $S=[0,1]\times[1/4,3/4]$.  In order
to upperbound the number of points that~$S$ contains, we subdivide~$S$
into eight horizontal strips, $S_1,\dots,S_8$, from top to bottom.  For
$i=1,\dots,8$, let~$a_i$
and~$b_i$ be (the y-coordinates of) the lower and upper boundaries
of~$S_i$.  We will fix~$a_i$ and~$b_i$ such that, for any point~$p$
in~$S_i$, each of $pl$, $pr$, and~$pt$ intersects either~$a_i$ or~$b_i$.

Observe that, for any point in~$S_i$, it holds that
its lower stubs intersect~$a_i$ if 
\begin{equation}
  \label{eq:upper}
  3/4 \cdot b_i \le a_i,
\end{equation}
whereas its upper stubs intersect~$b_i$ if 
\begin{equation}
  \label{eq:lower}
  a_i + (1-a_i)/4 = (3a_i+1)/4 \ge b_i.
\end{equation}
We may assume that~$S_i$ contains a point~$p_i$ on~$b_i$; otherwise we
simply let $p_i$ be the topmost point in~$S_i$ and restrict~$S_i$ to
the part between~$a_i$ and~$p_i$.  Let
\begin{equation}
  c_i = 3/4 \cdot b_i
  \label{EQ:c}
\end{equation} 
be the y-coordinate where the lower stubs of points on~$b_i$ end.  We
identify~$c_i$ with the line $y=c_i$.  For any point~$p$ in~$S_i$,
let~$I_p$ be the part of~$c_i$ delimited by the lower stubs
of~$p$.  Observe that, for $p,q \in S_i$ with $q \ne p$, it holds
that~$I_p$ and~$I_q$ are disjoint.  This is due to the fact that the
upper stubs of~$p$ and~$q$ both intersect~$b_i$.  Let $\delta_p$ be
the length of~$I_p$.  We say that $p$ \emph{consumes}~$\delta_p$.  By
the intercept theorem, we obtain that~$p$ consumes $\delta_p/1 \geq
(a_i-c_i)/a_i$ (which is what a point on~$a_i$ would consume).
The point~$p_i$ on~$b_i$ consumes $\delta_{p_i}/1 = (b_i-c_i)/b_i =
1/4$; hence, the other points together can consume at most~$3/4$.

To show that~$S_i$ contains at most five points besides~$p_i$,
we choose~$a_i$ and~$b_i$ such that
\begin{equation}
  6 \cdot \frac{a_i-c_i}{a_i} \geq 3/4.
  \label{EQ:less6} 
\end{equation}
(The reason for allowing equality is that the left or right boundaries
of~$S_i$ do not contain any points and hence, none of the intervals
on~$c_i$ intersects the boundary of~$R_t$.)

Combining Eqs.~\ref{EQ:c} and~\ref{EQ:less6} yields $a_i \ge 6/7 \cdot
b_i$.  This automatically implies Eq.~\ref{eq:upper}, so the
upper stubs of all points in~$S_i$ intersect~$b_i$.  Additionally, we
require Eq.~\ref{eq:lower} ($a_i \ge (4b_i-1)/3$) to make sure
that the lower stubs of all points in~$S_i$ intersect~$a_i$.

Now we can fix the upper and lower boundaries of the strips.  We start
with $b_1=3/4$ and then repeatedly set~$a_i$ to the tighter of the two
lower bounds (which, for $a_i \le 3/5$ is always the first one).  This
yields the following values:
\[
\begin{array}{lcccl}
  a_1 & = & 2/3       & = & b_2,\\ 
  a_2 & = & 4/7       & = & b_3,\\ 
  a_3 & = & 4/7 \cdot 6/7   & = & b_4,\\ 
  a_4 & = & 4/7 \cdot 6^2/7^2 & = & b_5,\\ 
  a_5 & = & 4/7 \cdot 6^3/7^3 & = & b_6,\\ 
  a_6 & = & 4/7 \cdot 6^4/7^4 & = & b_7,\\ 
  a_7 & = & 4/7 \cdot 6^5/7^5 & = & b_8, \text{ and}\\ 
  a_8 & = & 1/4.& & 
\end{array}
\]

So, each of the eight strips contains at most six vertices.  As it
turns out, we can tighten the analysis for~$S_1$ and~$S_8$.  Each
point in $S_1$ consumes at least $(a_1-c_1)/a_1 = 5/32$.  As above,
all such points (except~$p_1$) together must consume less than~$3/4$.
Therefore, $S_1$ contains at most five points.  Each point in~$S_8$
consumes $(a_8-c_8)/a_8 = 24337/7^6 \approx 0.21$; hence, $S_8$
contains at most four points.

Let's summarize.

\begin{lemma}
  The middle strip~$S$ contains at most 45 points.
\end{lemma}

\paragraph{The middle part of the bottom strip.} 

We consider the rectangle~$B = [1/4,3/4] \times [0,1/4]$ of length~$1/2$ and
height~$1/4$ between the cells~$C_l$ and~$C_r$. Similarly as for the middle
strip, we construct five cells $B_1$ to $B_5$ such that all stubs to the
extreme points $r, l$ and $t$ cross the cell boundaries.
We denote the left and right boundaries of cell $B_i$ by $a_i$
and~$b_i$ and set $a_1 = 1/4$, $b_1=a_2=1/3$, $b_2=a_3=4/9$, 
$b_3=a_4=5/9$, $b_4=a_5=2/3$, $b_5=3/4$.  Clearly, for every point $p$
in any cell $B_i$, it holds that the stubs $pl$ and $pr$ intersect the
left and right boundaries $a_i$ and $b_i$, respectively. For the upper
stub~$pt$, we only know that it crosses the horizontal line $y=1/4$,
but not necessarily the upper boundary of~$B_i$.  We say that a
point~$p$ is a \emph{medial} point in cell $B_i$ if its stub~$pt$ does
intersect the upper boundary of~$B_i$.  We observe that no two medial
points can lie in the same cell~$B_i$ without causing stub
intersections.  Thus, for another point~$q$ in~$B_i$, the stub~$qt$ must
intersect either~$a_i$ or~$b_i$.  We call such a point a \emph{lateral}
point. 

\begin{figure}[tb]
 \centering
    \includegraphics{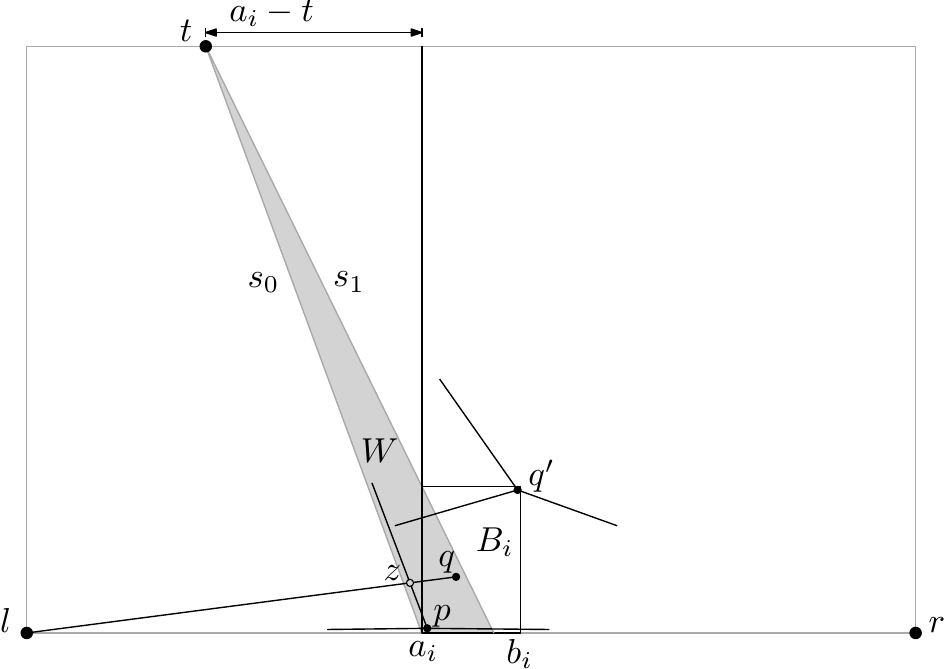}
    \caption{A cell $B_i$ with a medial point $q'$ and a lateral point $p$ whose stubs do not intersect. The two lateral points $p$ and $q$ cannot both exist since the stubs $pt$ and $ql$ must intersect. (The vertical axis in this figure is scaled by $2/3$.)}
    \label{fig:lower-middle} 
\end{figure}

In the following, we show that there can be at most one lateral point
in any cell $B_i$.  Without loss of generality, we can assume that $t
< a_i$. We consider the two rays~$s_0$ and~$s_1$ from~$t$ through the
two corners $(a_i,0)$ and $(a_i,1/4)$ of $B_i$, see
Fig.~\ref{fig:lower-middle}. These two rays define a wedge $W$. Let
$p$ and $q$ be two lateral points in $B_i$; then $p$ and $q$ must lie
in $B_i \cap W$. Let $p$ be the point whose ray from $t$ is left of
the ray from $t$ through $q$. Then the two line segments $\overline{ql}$
and $\overline{pt}$ must intersect in a point $z$ (otherwise the stubs
$pr$ and $qt$ would necessarily cross). Let $\delta_q =
|\overline{qz}|/|\overline{zl}|$. To avoid a crossing between the
stubs~$pt$ and~$ql$, we need that $|qz| \ge 1/4 (|qz| + |zl|)$, or
equivalently, that $\delta_q \ge 1/3$. 

Using the intercept theorem, we observe that~$\delta_q$ is maximized
if $p$ lies on $s_0$, $q$ lies on $s_1$, and they both lie on the
x-axis.  We apply the intercept theorem once more for the line $x=a_i$
and the line supported by $s_1$ to show that in this case
$|\overline{qz}| = (a_i-t)/3$.  Using $|\overline{zl}| = a_i$, we get
$\delta_q = 1/3 \cdot (1 - t/a_i) < 1/3$.  This contradicts $\delta_q
\ge 1/3$.  Thus, each cell~$B_i$ contains at most one lateral and
at most one medial point.

We summarize.

\begin{lemma}
The lower rectangle $B$ contains at most 10 points.
\end{lemma}

\paragraph{The left and the right part of the upper strip.}

In the following, we consider the rectangles $L=[0,3t/4] \times
[3/4,1]$ and $R=[(3t+1)/4,1]\times[3/4,1]$, separated by the upper
central square~$C_t$, which has size $1/4 \times 1/4$, is
adjacent to~$t$, and is defined by the stubs~$tl$ and~$tr$.
Recall that we assume $0 < t \leq 1/2$. 

We subdivide~$R$ into five height-$1/4$ rectangles~$R_1,\dots,R_5$,
from left to right, and analyze how many points each rectangle can
contain at most.  The analysis for~$L$ is symmetric.

Note that the two lower stubs of any point in~$R$ intersects
the horizontal line $y=3/4$.
To make sure that the upper stub of any point in cell $R_i$ intersects
the left boundary with x-coordinate~$a_i$ of~$R_i$, the
x-coordinate~$b_i$ of the right boundary of~$R_i$ has to fulfill
$(b_i-t)/4 \geq b_i-a_i$ and, hence, $b_i \leq (4a_i-t)/3$.  (Note that we
assume that the right boundary of~$R_i$ is not part of~$R_i$.)  This
yields the following boundaries.
\[
\begin{array}{lcccl}
       && 3/4 \cdot t + 1/4 &=& a_1,\\            
       b_1& = &2/3\cdot t + 1/3 &= &a_2,\\          
       b_2& = &5/9 \cdot t + 4/9& = &a_3,\\        
       b_3&  = &11/27 \cdot t + 16/27& =&  a_4,\\     
       b_4&  = &17/81 \cdot t + 64/81& =&  a_5, \text{ and}\\ 
       b_5&  =&  1.& & 
\end{array}
\]

Observe that the lower stubs of
all points in the same cell have to be nested: Let $p$ be a point
in a cell $R_i$ and consider the line $s$ through $l$ and $p$.  Assume
there is a point $q$ above $p$ such that the lower stubs of $p$ and
$q$ are not nested.  Then $q$ has to be to the right of $s$.  However,
by the way the width of $R_i$ is constructed, the left stub of $q$
intersects the vertical line~$a_i$. So, the left stub of $q$
intersects the right stub of~$p$.

Now we analyze how many points can be stacked on top of each other in
each subrectangle, depending on its width, its distance to $t$, and
the \emph{$l$-shadow} of the stub~$tr$, i.e., the set of all
points~$p$ such that the stubs $tr$ and $pl$ intersect.

Consider first $R_1$.  Observe that the left stub of any point~$p$
in~$R_1$ must leave~$R_1$ through its bottom edge.  Otherwise~$p$
would lie in the $l$-shadow of $tr$.  Hence, there are no two points
with nested lower stubs in $R_1$.  Otherwise the left stub of the
upper point would intersect the upper stub of the lower point.  Thus,
$R_1$ contains at most one point.

For the remaining four subrectangles it is easy to see that they
contain at most five points each.  This yields to following lemma.

\begin{figure}[tb]
 \centering
    \includegraphics{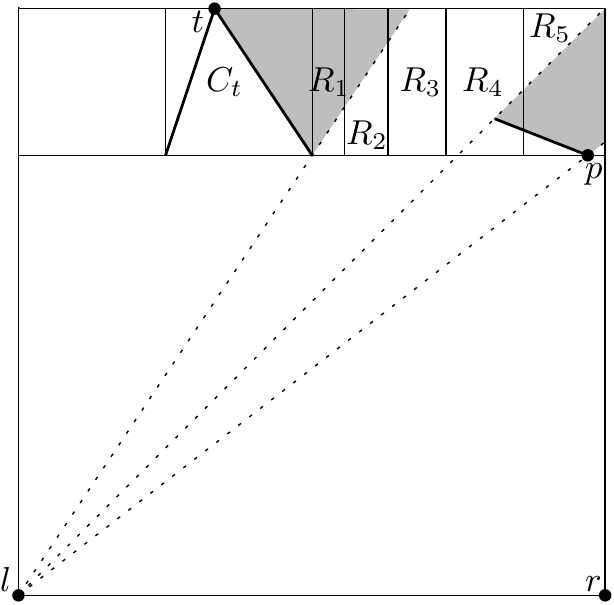}
    \caption{The lines from point~$l$ to the endpoints of the
      stubs~$tr$ and~$pt$ give rise to (gray) $l$-shadows where no point
      can be placed.}
    \label{fig:upper}
\end{figure}

\begin{lemma}
  The rectangles~$L$ and~$R$ each contain at most 21 points.
\end{lemma}

\paragraph{The $(1/4 \times 1/4)$-squares $C_l$, $C_r$, and $C_t$.}

Our approach for this part follows a suggestion of Ga{\v s}per
Fijav{\v z}.  We consider the square~$C_l$; for the two other
squares~$C_r$ and~$C_t$, we can argue analogously and get the same
bound.  Let $l, p_1,\dots,p_k$ be the set of points contained
in~$C_l$.

First, we observe that the stubs from $p_1,\dots,p_k$ to $t$ and $r$
intersect the upper and right boundary of~$C_l$.  Hence, the points
$p_1,\dots,p_k$ together with their stubs to~$r$ and~$t$ form a nested
structure.  This means that we can order the points such that, for
$i=2,\dots,k$, the point~$p_i$ lies between the stubs of $p_{i-1}$,
the point $p_k$ being innermost.  Now we define $\alpha_1,
\alpha_2,\dots, \alpha_k$ to be the angles at point $r$ formed by the
lines $\overline{rl}$ and $\overline{rp_i}$.  Analogously, we have
angles $\beta_1,\dots,\beta_k$ at point $t$.  We consider only the
angles of type~$\alpha_i$.  Analogous observations hold for the angles
of type~$\beta_i$, and the resulting bounds are the same.

From the nesting, we see that the sequence $\alpha_i$, $i=1,\dots,k$ is
monotonously increasing.  Even stronger, we have the following claim.


\begin{claim}
  For $1 < i \leq k$ it holds that $\alpha_i \geq 1.3 \cdot \alpha_{i-1}$.
\end{claim}

\begin{figure}
  \begin{minipage}[b]{.5\textwidth}
    \centering
    \includegraphics[width=\textwidth]{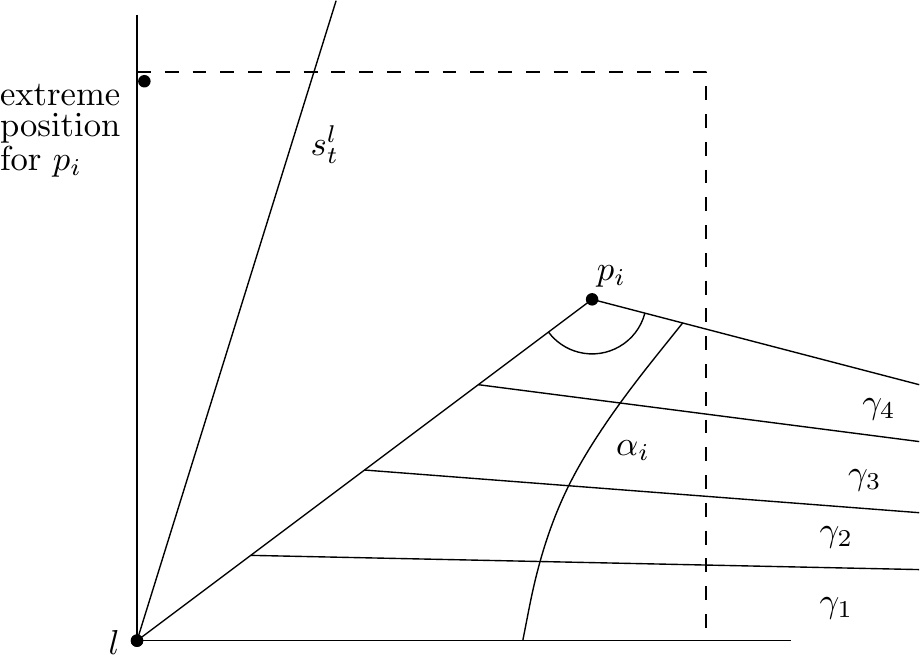}
    \caption{Angles increase by a factor of at least $1.3$.}
    \label{fig:angles}
  \end{minipage}
  \hfill
  \begin{minipage}[b]{.46\textwidth}
    \centering
    \includegraphics[width=\textwidth]{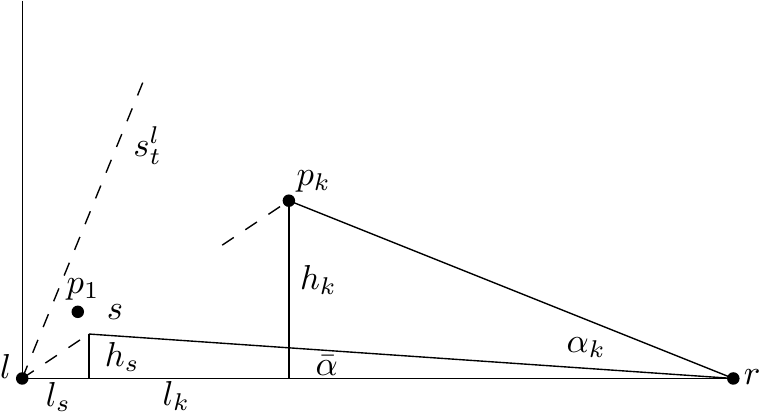}
    \caption{Ratio between the smallest and the largest angle.}
    \label{fig:ratio}
  \end{minipage}
\end{figure}

\begin{pf}
  Consider the segment from point $l$ to $p_i$.  We subdivide it into
  four segments of equal length; see Fig.~\ref{fig:angles}.  This
  defines the four angles $\gamma_1,\dots,\gamma_4$ by the connecting
  lines from point~$r$.  We have $\alpha_i = \gamma_1 + \dots +
  \gamma_4$ and $\alpha_{i-1} \leq \gamma_1 + \gamma_2 + \gamma_3 =:
  \gamma$.  It remains to prove that $\alpha_i \geq 1.3 \gamma$.

  The ratio of $\gamma_4$ and $\gamma$ and, hence, the ratio of
  $\alpha_i$ and $\gamma$ is smallest if the angle $\angle r,p_i,l$ is
  minimized, i.e., if $p_i$ lies in the upper left corner of
  $C_l$. Hence, $\alpha_i / \gamma \geq \arctan(1/4) / \arctan(3/16) >
  1.3$. Note that in general $p_i$ must be to the right of stub $lt$,
  so the ratio is even slightly better.
\end{pf}


Next, we restrict the range of the smallest and largest angle.  Then
we can easily compute the number $k$ of points.  

Let $s$ be the endpoint of the stub $l{p_k}$.  Consider the two lines
$\overline{ts}$ and $\overline{rs}$.  The point~$p_1$, which defines
the angle $\alpha_1$ and $\beta_1$ respectively, has to lie either
above $\overline{rs}$ or to the right of $\overline{ts}$ or both. We
assume, without loss of generality, the first case, so the angle
formed by $l,r,s := \bar{\alpha} \leq \alpha_1$.  We call the length
of the base line, which is the distance between $l$ and $r$, to be $d
= 1$.

We compute $\tan(\bar{\alpha}) = h_s/d-l_s = (h_k/4)/d-l_s$ and
$\tan(\alpha_k) = h_k/(d-4l_s) = h_k/(d-4l_s)$; see
\figurename~\ref{fig:ratio}, where $h_k$ and~$h_s$ are the minimum
distances of~$p_k$ and~$s$, respectively, to the base
line~$\overline{lr}$.  This yields the ratio

\[\frac{\tan \bar{\alpha}}{\tan \alpha_k} =
\frac{h_k}{4(d-l_s)}\frac{d-4l_s}{h_k} = 1 - \frac{3d}{4(d-l_s)} = 1 -
\frac{3}{4(1 - l_s/d)}.\] 
Using $l_s \leq (1/4)^2$, this yields
$\tan \bar{\alpha} / \tan \alpha_k \geq 1/5$. 
From the Taylor seriesexpansion of the tangent function we know that
$\tan \alpha > \alpha$, for all $0 < \alpha < \pi/2$, in particular
$\tan \alpha = c_\alpha \alpha$ with $c_\alpha > 1$ monotonically
increasing with $\alpha$. 

Hence, we conclude that \[\frac{\alpha_1}{\alpha_k} =
\frac{c_{\alpha_k}\tan \alpha_1}{c_{\alpha_1} \tan \alpha_k} >
\frac{\tan \bar\alpha}{\tan \alpha_k} \geq 1/5.\] 
This yields $\alpha_k \geq (1.3)^{k-1}\cdot \alpha_1 > (1.3)^{k-1}
\cdot 1/5 \cdot \alpha_k$, which in turn implies $k < \log
  5 / \log 1.3 + 1 \leq 6.2$. 

If~$p_1$ lies to the right of $\overline{ts}$, we analogously obtain
\[\frac{\beta_1}{\beta_k} > 1 - \frac{3}{4(1-l_s/d)}\]
where $d>1$ is the distance of~$l$ and~$t$ and $l_s \leq 1/4 \cdot
\sqrt{2}/4$ is the length of the projection of the segment
$\overline{ls}$ to the line $lt$, hence $\beta_1 > 1/6 \cdot \beta_k$.

Arguing along the lines of the first case, we get $\beta_k >
(1.3)^{k-1} \cdot 1/6 \cdot \beta_k$, and derive $k < 7.9$.
  
\begin{lemma}
  The squares $C_l$, $C_t$, and~$C_r$ each contain at most eight points. 
\end{lemma}

This finishes the proof of Theorem~\ref{thm:improved}.

\section{Improved Bounds for Specific Graph Classes}
\label{sec:specific-classes}

In this section, we improve, for specific graph classes, the result of
Bruckdorfer and Kaufmann \cite{bk-mecbe-FUN12} which says that $K_n$
(and thus, any $n$-vertex graph) has a $1/\sqrt{4n/\pi}$-SHPED.  
In other words, $K_n$ has a $\delta$-SHPED if $n \le \pi/(4\delta^2)$.
We give two constructions for complete bipartite graphs and one for
graphs of bounded bandwidth.

\paragraph{Complete Bipartite Graphs.} 

\begin{figure}[tb]
  \centering
  \subfloat[\label{fig:Knn}$1/4$-SHPED of $K_{8,8}$]%
  {\includegraphics[height=3.5cm,bb=75 240 440 560,clip=true]{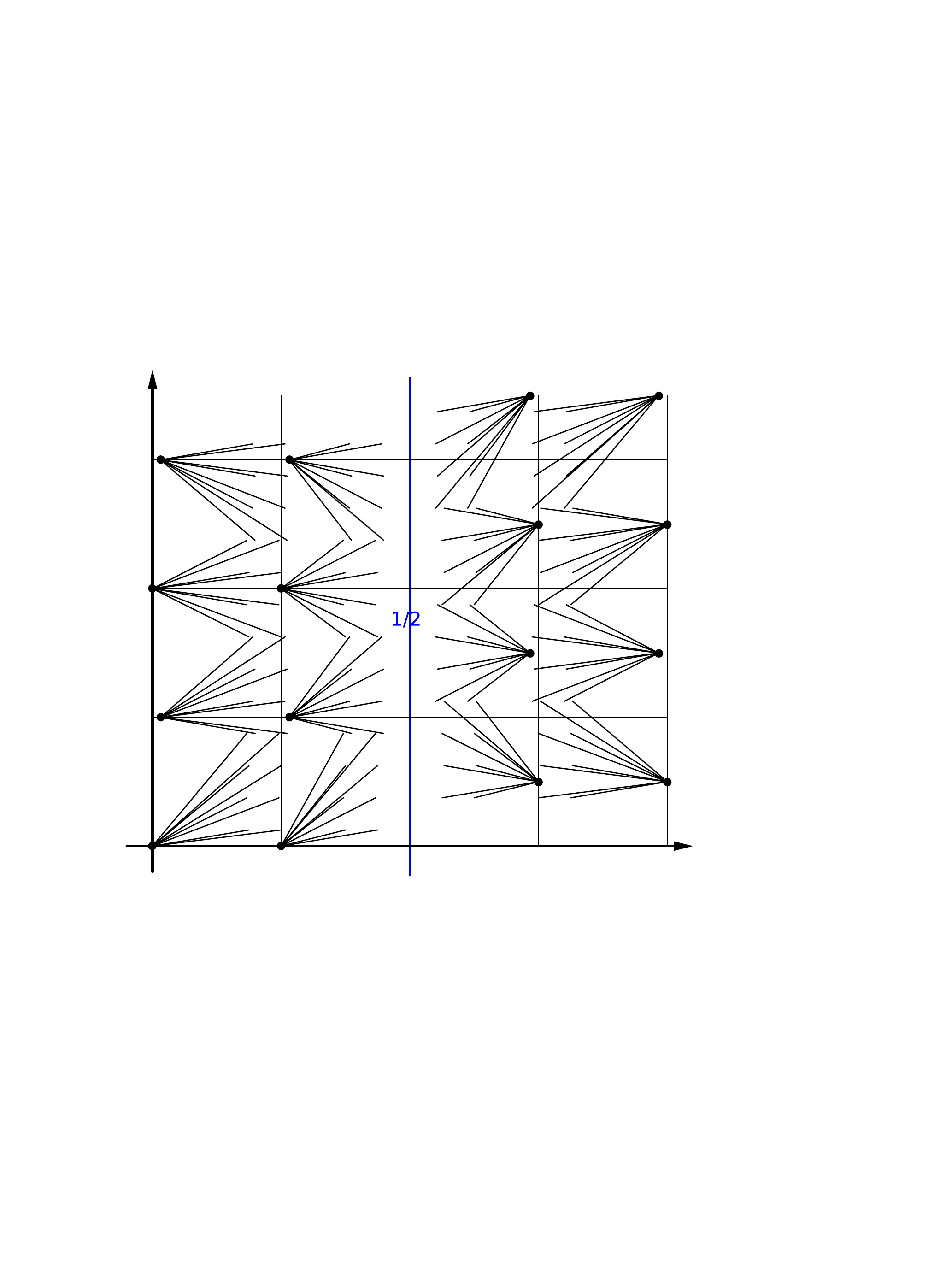}} 
  \hfil
  \subfloat[\label{fig:Kkn}$1/4$-SHPED of $K_{8,9}$]%
  {\includegraphics[height=3.5cm,bb=30 260 565 540,clip=true]{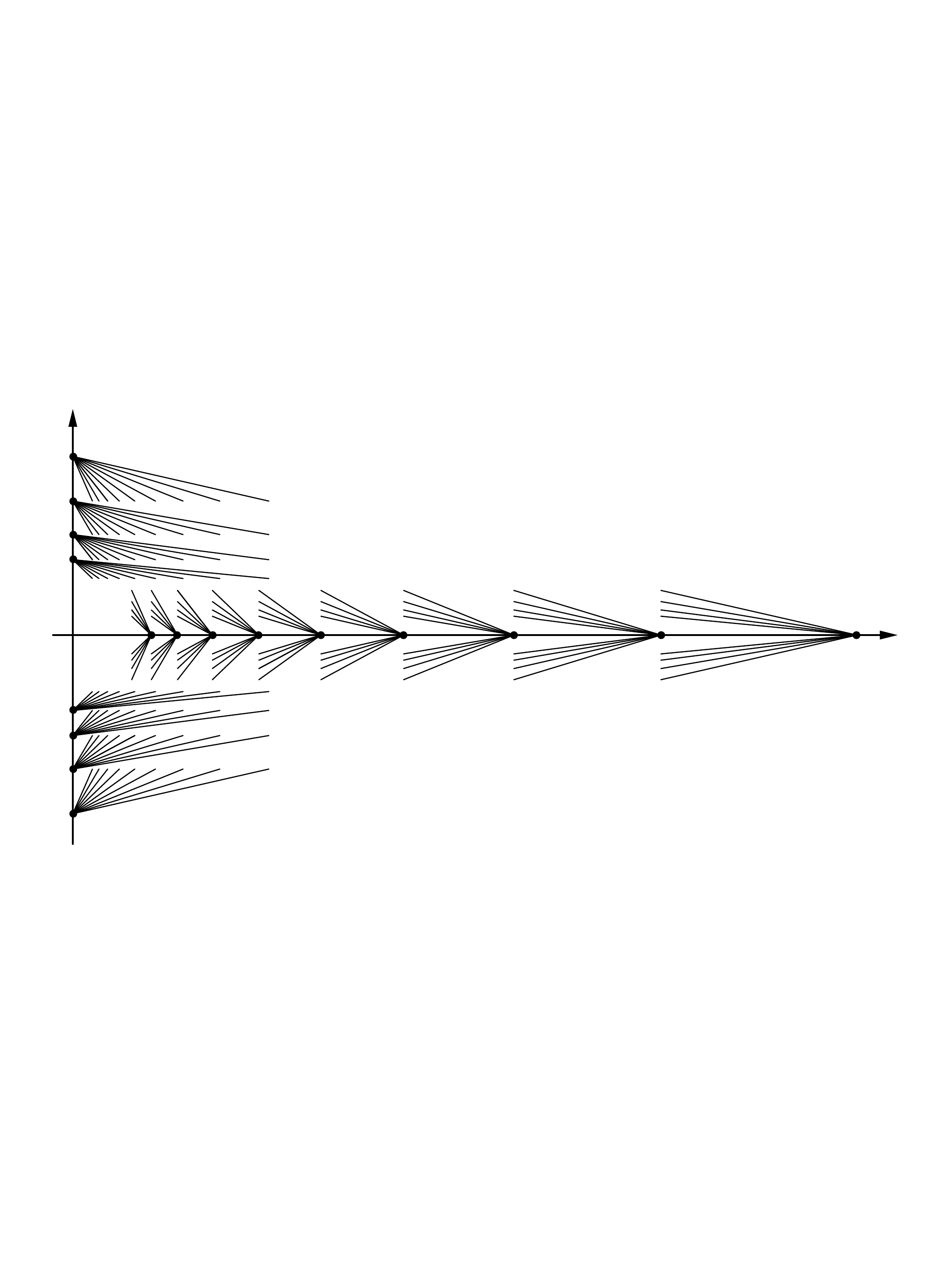}}

  \caption{Two methods for drawing complete bipartite graphs as SHPEDs.}
  \label{fig:bipartite}
\end{figure}

Our first construction is especially suitable if
both sides of the bipartition have about the same size. The drawing is
illustrated in \figurename~\ref{fig:Knn}. Note that in the figure the x- and 
y-axes are scaled differently. 

\begin{theorem}
  \label{thm:Knn}
  The complete bipartite graph $K_{n,n}$ has a $\delta$-SHPED if 
  \[
    n \leq
    \left\lfloor \frac{1}{\delta} \right\rfloor 
    \cdot 
    \left\lfloor\left\lfloor \frac{\log 1/2}{\log(1-\delta)} 
      \right\rfloor\right\rfloor,
  \] 
  where $\lfloor\lfloor r\rfloor\rfloor$ denotes the
  largest integer that is strictly less than~$r$.
\end{theorem}
\begin{pf}
  Let $k=\left\lfloor \frac{1}{\delta} \right\rfloor$
  and $\ell = \left\lfloor\left\lfloor \frac{\log 1/2}{\log(1-\delta)} 
  \right\rfloor\right\rfloor$. 
  The latter implies that $(1-\delta)^\ell > \frac{1}{2}$.

  Divide the plane at the vertical line $x = 1/2$ into two half planes, one for each side of the bipartition,
  to which we will refer as the right-hand side and the left-hand
  side. In each half plane draw the $n$ vertices on a (perturbed) $k \times \ell$
  grid.  More precisely, 
  for a horizontal line, let $\epsilon \geq 0$ such that $(1-\delta)^\ell > \frac{1}{2} + \epsilon$. Draw the vertices with
  x-coordinates
  \[
     (1-\delta)^i - \epsilon \text{ and } 1 - (1-\delta)^i + \epsilon,i=0,\dots,\ell-1. 
  \] 
  Draw the vertices on the left-hand side with y-coordinates
  $0,\dots,k-1$ and the vertices on the right-hand side with
  y-coordinates $0+\sigma,\dots,k-1+\sigma$ where $0 < \sigma < 1$ is
  chosen such that no two vertices on the right-hand side are
  collinear with a vertex on the left-hand side and vice versa. All
  edges are between a vertex on the left-hand side and a vertex on the
  right-hand side. 
 
  Then for any two vertices the bounding boxes of their
  incident stubs are disjoint up to their boundaries. Intersections of
  the stubs on the boundaries can be avoided by a suitable choice of
  $\epsilon$.  
  \begin{enumerate}
  \item
    If $v$ is a vertex on the right-hand side with x-coordinate
    $(1-\delta)^i - \epsilon$. Then the projection to the x-axis of the
    longest edge incident to $v$ has length
    $(1-\delta)^i - \epsilon$. Hence all stubs incident to $v$ are in
    the vertical strip bounded by $x = (1-\delta)^i - \epsilon$ and $x
    = (1-\delta)^i -\epsilon - \delta((1-\delta)^i-\epsilon) \geq
    (1-\delta)^{i+1} - \epsilon > 1/2$. 
    The latter inequation follows since $i+1 \leq \ell$. 
  \item
    Let $v_i$ be a vertex with y-coordinate $\sigma + i$,
    $i=0,\dots,k-1$.  Then the projection to the y-axis of the longest
    edge incident to $v_i$ and above $v$ has length $k-1-i-\sigma$ while
    the projection to the y-axis of the longest edge incident to
    $v_i$ and below $v$ has length $i+\sigma$. Hence the projection
    to the y-axis of the stubs incident to $v_i$ and $v_{i+1}$ do
    not intersect if $\delta(k-1-i-\sigma) + \delta(i+1+\sigma) < 1$
    which is fulfilled if $k < 1/\delta$. 
    If $k = 1/\delta$ then draw the vertices on the horizontal lines
    $y = i$ and $y = i + \sigma$ for even $i$ with $\epsilon =0$ and
    the vertices on the other horizontal lines with a slightly
    positive $\epsilon$ such that the end points of the stubs do not intersect.
  \end{enumerate}
  A symmetric argument holds for the vertices on the left-hand side.
\end{pf}

Our second construction is especially suitable if one side of the
bipartition is much larger than the other. The drawing is illustrated
in \figurename~\ref{fig:Kkn}.
\begin{theorem}
  \label{thm:Kkn}
  For any integers $n>0$ and $k < {\log \delta}/{\log(1 - \delta)}$,
  the complete bipartite graph $K_{2k,n}$ has a $\delta$-SHPED.
\end{theorem}
\begin{pf}
  Draw the $n$ vertices on the x-axis with x-coordinate
  $
     x_i = 1/(1-\delta)^{i-1}, 
     i = 1,\dots,n
  $
  and the $2k$ vertices on the y-axis with y-coordinate
  $
     y_i = 1/(1-\delta)^{i-1}, 
     i = 1,\dots,k
  $
  and $-y_i, i = 1,\dots,k$.
  All edges are between a vertex on the y-axis and a vertex on
  the x-axis. To show that no stubs intersect, we establish the
  following two properties on the regions that contain the stubs.
  \begin{enumerate}
  \item The stubs incident to $(0,\pm y_i),i=2,\dots,k$ are in the
    horizontal strip bounded by $y=\pm y_i$ and $y= \pm y_{i-1}$:

    The projection to the y-axis of any stub incident to $(0,y_i)$ has
    length $\delta \cdot y_i$, hence it stops at $y = (1-\delta) \cdot
    \left(\frac{1}{1-\delta}\right)^{i-1} =
    \left(\frac{1}{1-\delta}\right)^{i-2} = y_{i-1}$.
  \item The stubs incident to $(0,x_i),i=2,\dots,n$ are in the
    rectangle bounded by $y = \pm (1-\delta)$, $x = x_{i-1}$, and $x =
    x_{i}$ (where $x_0 = 1 - \delta$):

    As above, the projection of any stub incident to $(0,x_i)$ stops
    at $x= x_{i-1}$.  The absolute value of the projection to the
    y-axis is bounded by $\delta \cdot y_k = \delta \cdot
    \left(\frac{1}{1-\delta}\right)^{k-1}$ which is less than
    $1-\delta$ if $k < {\log \delta}/{\log(1 - \delta)}$.
  \end{enumerate}
  Since the stubs incident to $(0,\pm y_1)$ lie in the horizontal strip
  bounded by $y=\pm 1$ and $y=\pm (1-\delta)$, it follows that any two
  stubs are disjoint.
\end{pf}

\paragraph{Graphs of Bounded Bandwidth.} 

Recall that the \emph{$k$-circulant} graph $C_n^k$ with $n$ vertices
and $0 \le k < n$ is the undirected simple graph whose vertex set is
$\{v_0,\dots,v_{n-1}\}$ and whose edge set is $\{v_iv_j \colon |j-i| \le
k\}$.  When we specify the index of a vertex, we implicitly assume
calculation modulo~$n$.  Note that $C_n^1 = C_n$ and $C_n^{n/2} =
K_n$.
We provide $\delta$-SHPED constructions for $k$-circulant and
bandwidth-$k$ graphs where $\delta=\Theta\big(1/\sqrt{k}\,\big)$.  For ease of
presentation, we assume that $\sqrt{k}$ and $n\big/\sqrt{k}$ are integers.

\begin{figure}[tb]
  \centering
  \subfloat[path case\label{sfg:path}]{\includegraphics{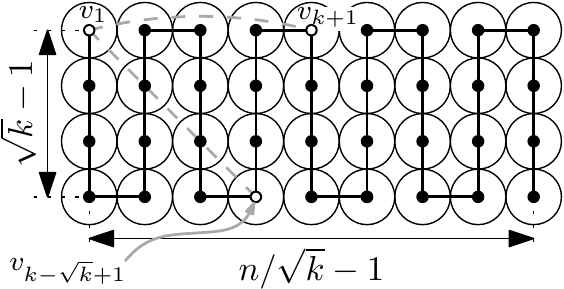}}
  \hfil
  \subfloat[circle case\label{sfg:circle}]{\includegraphics{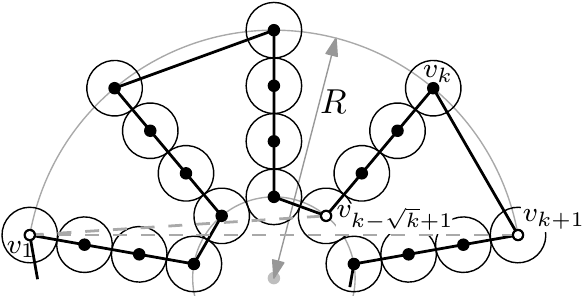}}
  \caption{SHPEDs for bandwidth-$k$ and $k$-circulant graphs.} 
  \label{fig:k-path}
\end{figure}

First, let $G$ be a graph of bandwidth~$k$, i.e., the vertices of
$G$ can be ordered $v_1,\dots,v_n$ and for each edge $v_iv_j$ it holds
that $|j-i| \le k$.  We draw~$G$ as a $\delta$-SHPED as follows.  We
map the vertices of~$G$ to the vertices of an integer grid of
$\big(n\big/\sqrt{k} \times \sqrt{k}\,\big)$ points such that the sequence of
vertices $v_1,\dots,v_n$ traverses the grid column by column in a snake-like
fashion, see \figurename~\ref{sfg:path}.  

The distance from any vertex to its $k$-th successor is at most 
$\sqrt{\big(\sqrt{k}-1\big)^2 + k} < \sqrt{2k}$, see the
two dashed line segments in \figurename~\ref{sfg:path}.  Setting
$\delta = 1/\big(2\sqrt{2k}\,\big)$ ensures that each stub is contained in the
radius-$1/2$ disks centered at the vertex to which it is incident; see
\figurename~\ref{sfg:path}.  Since the disks are pairwise disjoint, so
are the stubs.

For the $k$-circulant graph~$C_n^k$, we modify this approach such that
start and end of the snake coincide.  In other words, we deform our
rectangular section of the integer grid into an annulus; see
\figurename~\ref{sfg:circle}.  We additionally assume that
$n/\sqrt{k}$ is even.

The inner circle circumscribes a regular
$\big(n/\sqrt{k}\,\big)$-gon~$\Pi$ of edge length~1.  

We place the vertices of~$C_n^k$ on rays
that go from the center of the annulus through the vertices of~$\Pi$.
On each ray, we place $\sqrt{k}$ vertices at distance~1 from one
another, starting from the inner circle and ending at the outer
circle.  The sequence again traverses the stacks of vertices in a
snake-like fashion.

A vertex $v$ can be reached from its $j$-th ($j<k$) successor $s$, by
traversing at most $3\sqrt{k}-2$ segments of length 1: at most $\sqrt{k}-1$
segments from $s$ to the inner circle, at most $\sqrt{k}$ segments on the
inner circle, and at most $\sqrt{k}-1$ segments from the inner circle to
$v$. Hence, the maximum distance of two adjacent vertices is less than
$3\sqrt{k}$, and we can choose $\delta=1/(6\sqrt{k})$.



\begin{theorem}
  Let $2 \le k \le n$ and assume that $\sqrt{k}$ and $n\big/\sqrt{k}$ are
  integers.  Then any graph of bandwidth~$k$ has a
  $1\big/\big(2\sqrt{2k}\,\big)$-SHPED.  If additionally 
  $n\big/\sqrt{k}$ is even, the $k$-circulant graph~$C_n^k$ has a
  $1\big/\big(6\sqrt{k}\,\big)$-SHPED.
\end{theorem}

\section{Geometrically Embedded SPEDs}
\label{sec:speds}

Bruckdorfer and Kaufmann~\cite{bk-mecbe-FUN12} gave an integer-linear
program for \textsc{MaxSPED} and conjectured that the problem is
NP-hard.  Indeed, there is a simple reduction from \textsc{Planar3SAT} 
\cite{ks-pc-12}.  In this section, we first show that the problem can be
solved efficiently for the special case of graphs that admit a
2-planar geometric embedding.  Then we turn to the dual problem
\textsc{MinSPED} of minimizing the ink that has to be erased in order
to turn a given drawing into a SPED.

\subsection{Maximizing Ink in Drawings of 2-Planar Graphs}
\label{sub:two-planar}

In this section we prove that, given a 2-planar geometric embedding~$\Gamma$ 
of a 2-planar graph~$G$ with $n$ vertices, we can compute a maxSPED, i.e., a SPED that maximizes the total stub length,
in $O(n \log n)$ time.  Recall that a graph $G$ is 2-planar if it admits a 
simple drawing on the plane where each edge is crossed at most twice.

Given~$G$ and~$\Gamma$, we define a simple undirected graph $C$ 
as follows. $C$ has a vertex $v_e$ for each edge $e$ of $G$. Two
vertices $v_e$ and $v_{e'}$ of $C$ are connected by an edge if and
only if $e$ and $e'$ form a crossing in $\Gamma$. Such a graph is in
general non-connected. Furthermore, since the maximum degree of $C$ is
$2$, a connected component of $C$ is either a path (possibly formed by
only one edge) or a cycle. 

Let $C_i$ be a connected component of $C$.  We define a total ordering
of the vertices of $C_i$. Namely, if $C_i$ is a path such an ordering
is directly defined by the order of its vertices along the path
(rooted at an arbitrary end vertex). If $C_i$ is a cycle, we simply
delete an arbitrary edge of the cycle, obtaining again a path and the
related order.  That means, if we consider the subdrawing
$\Gamma_{i}$ of $\Gamma$ induced by the vertices of $C_i$ (edges
of $G_i$), such a drawing is formed by an ordered sequence of edges
(according to the ordering of the vertices of $C_i$),
$e_1,\dots,e_{n_i}$, such that $e_j$ crosses $e_{(j+1) \bmod n_i}$ for
$j=1,\dots,n_i-1$ in case of a path, and $j=1,\dots,n_i$ in case of a
cycle.

\begin{wrapfigure}[8]{r}{31ex}
  \vspace{-3ex}
  \hfill\includegraphics{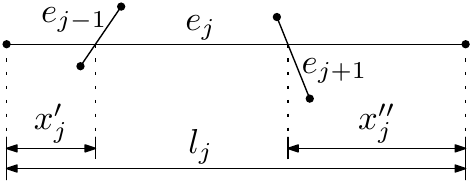}
  \caption{Notation used in the DP.}
  \label{fi:2-planar-mSPED}
\end{wrapfigure}

We will use the following notation: $l_j$ is the total length of the
edge $e_j$; $x_j'$ is the length of the shortest stub of $e_j$ defined by the
crossing between $e_{j-1}$ and $e_j$, called the backward stub;
$x_j''$ is the length of the shortest stub of $e_j$ defined by the crossing
between $e_j$ and $e_{j+1}$, called the forward stub. See also
\figurename~\ref{fi:2-planar-mSPED}.

Consider now the subdrawing $\Gamma_{i}$, and assume that
$e_1,\dots,e_{n_i}$ form a path in $C_i$.  If $n_i = 2$, the 
maximum total length of the stubs is
$\kopt= \max\{l_1+2x_2',l_2 + 2x_1''\}$.

\begin{figure}[tb]
  \centering
  \subfloat[][]{\label{fi:2-planar-mSPED-ex-a}%
    \includegraphics[scale=.9]{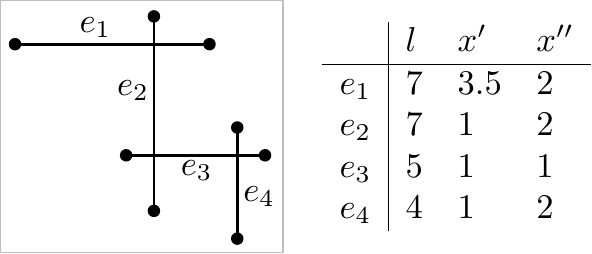}}
  \hfil
  \subfloat[][]{\label{fi:2-planar-mSPED-ex-b}%
    \includegraphics[scale=.9]{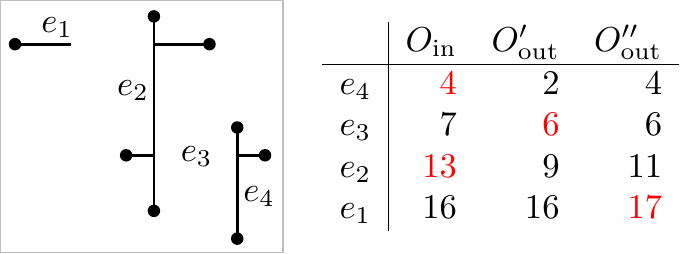}}
  \caption{(a) A 2-planar drawing $\Gamma$ and (b) a maxSPED 
    of $\Gamma$ computed by the DP algorithm.}
  \label{fi:2-planar-mSPED-ex}
\end{figure}

In the general case, we can process the path edge by edge, having at
most three choices for each edge: $(i)$ we can draw it entirely,
$(ii)$ we can draw only its backward stubs, or $(iii)$ we can draw
only its forward stubs. The number of choices we have at any step is
influenced only by the previous step, while the best choice is
determined only by the rest of the path. Following this approach, let
$\gamma_{i}$ be a maxSPED for $\Gamma_{i}$ and consider the 
choice done for the first edge $e_1$ of the path. The total length of the 
stubs in $\gamma_{i}$, minus the length of the stubs assigned to $e_1$, 
represents an optimal solution for $\Gamma_i \setminus e_1$, under the initial 
condition  defined by the first step, otherwise, $\gamma_{i}$ could be 
improved, a contradiction. I.e., the optimality principle holds
for our problem. Thus, we can exploit the following dynamic
programming (DP) formulation, where $\Oin(e_j)$ describes the maximum total 
length of the stubs of $e_j,\dots,e_{n_i}$ under the choice~$(i)$ for $e_j$, $\Oout'(e_j)$ describes the choice $(ii)$ and
$\Oout''(e_j)$ describes the choice~$(iii)$.
\begin{subequations}\label{eq:2plan-sped}
\begin{align}
\Oin(e_j) &=
\begin{cases} l_j + \max\{\Oout'(e_{j+1}), \Oout''(e_{j+1})\} &\text{if $x_{j+1}' \geq x_{j+1}''$,}
\\
l_j + \Oout'(e_{j+1}) &\text{if $x_{j+1}' < x_{j+1}''$.}
\end{cases}\label{eq:2plan-sped-oin}
\\
\Oout'(e_j) &= 
\begin{cases} 2x_j' + \max\{\Oout'(e_{j+1}),\Oout''(e_{j+1})\} & \text{if $x_j' > x_j''$ and $x_{j+1}' \geq x_{j+1}''$,}
\\
2x_j' + \Oout'(e_{j+1}) & \text{if $x_j' > x_j''$ and $x_{j+1}' < x_{j+1}''$,}
\\
\multicolumn{2}{l}%
{2x_j' + \max\{\Oin(e_{j+1}), \Oout'(e_{j+1}), \Oout''(e_{j+1})\}  
\qquad \quad \; \; \; \text{if $x_j' \leq x_j''$.}}
\end{cases}\label{eq:2plan-sped-oout1}
\\
\Oout''(e_j) &= 2x_j'' + \max\{\Oin(e_{j+1}), \Oout'(e_{j+1}), \Oout''(e_{j+1})\}\label{eq:2plan-sped-oout2}
\end{align}
\end{subequations}
In case of a path, we store in a table the values of $\Oin(e_j)$,
$\Oout'(e_j)$ and $\Oout''(e_j)$, for $j=1,\dots,n_i$, through a
bottom-up visit of the path (from $e_{n_i}$ to $e_1$). Since $e_1$ and $e_{n_i}$ do not cross, we have $x_1' = l_1/2$
and $x_{n_i}'' = l_{n_i}/2$. Then, the maximal value of ink is
given by $\kopt = \max\{\Oin(e_1), \Oout'(e_1),
\Oout''(e_1)\}$. See \figurename~\ref{fi:2-planar-mSPED-ex} for an
example.

In case of a cycle, we have that $e_1$ and $e_{n_i}$ cross each other,
thus, in order to compute the table of values we must assume an
initial condition for $e_{n_i}$. Namely, we perform the bottom-up
visit from $e_{n_i}$ to $e_1$ three times. The first time we consider
as initial condition that $e_{n_i}$ is entirely drawn (choice
$\Oin(e_{n_i})$), the second time we consider only the backward
stubs drawn (choice $\Oout'(e_{n_i})$), and the third time we
consider only the forward stubs drawn (choice
$\Oout''(e_{n_i})$). Every initial condition will lead to a table
where, in general, we do not have all the three possible choices for
$e_1$ (i.e., some choices are forbidden due to the initial
condition). Performing the algorithm for every possible initial
condition and choosing the best value yields the
optimal solution \kopt.
The algorithm described above leads to the following result.

\begin{theorem}
  Let $G$ be a graph with $n$ vertices, and let $\Gamma$ be a
  2-planar geometric embedding of $G$.  A maxSPED of $\Gamma$ can
  be computed in $O(n\log n)$ time.
\end{theorem}

\begin{pf}
  Consider the above described algorithm, based on the 
  DP formulation defined by the set of equations~(\ref{eq:2plan-sped}). We
  already showed how this algorithm computes a maxSPED of
  $\Gamma$. The construction of the graph $C$ requires time $O(m
  \log m)$ with a standard sweep-line algorithm for computing the $O(m)$ line-segment intersections~\cite{bo-arcgi-79}. Once $C$ has been constructed,
  ordering its vertices requires $O(n_C)$ time, where $n_C \in O(m)$
  is the number of vertices of $C$. Performing a
  bottom-up visit and up to three top-down visits of every path or
  cycle takes $O(m)$ time. Thus, the overall time complexity is $O(n \log
  n)$, since for 2-planar graphs $m \in O(n)$~\cite{pt-gdwfc-97}.
\end{pf}

We finally observe that the restricted $0/1$-\textsc{MaxSPED} problem 
for 2-planar drawings, where each edge is either drawn or erased completely, 
may be solved through a different approach. Indeed, we can exploit a
maximum-weight SAT formulation in the \textsc{CNF+($\le$2)} model,
where each variable can appear at most twice and only with positive
values~\cite{ps-avw2+-07}.  
Roughly speaking, we map each edge to a variable, with 
the weight of the variable equal to the length of its edge, and define
a clause for each crossing.  Applying an algorithm of Porschen and
Speckenmeyer~\cite{ps-avw2+-07} for \textsc{CNF+($\le$2)}
solves $0/1$-\textsc{MaxSPED} in $O(n^3)$ time.  However, our
algorithm solves a more general problem in less time. 

\subsection{Erasing Ink in Arbitrary Graph Drawings}
\label{sub:erasing}


In this section, we consider the problem \textsc{MinSPED}, which is dual to
\textsc{MaxSPED}.  In \textsc{MinSPED}, we are given a graph with a
straight-line drawing (i.e., a geometric graph), and the task is to
erase as little of the edges as possible in order to make it a SPED.

We will exploit a connection between the NP-hard minimum-weight 2-SAT
problem (\textsc{MinW2Sat}) and \textsc{MinSPED}.  Recall that
\textsc{MinW2Sat}, given a 2-SAT formula, asks for a satisfying
variable assignment that minimizes the total weight of the true
variables.  There is a 2-approximation algorithm for \textsc{MinW2Sat}
that runs in $O(vc)$ time and uses $O(c)$ space, where $v$ is the
number of variables and $c$ is the number of clauses of the given
2-SAT formula~\cite{br-eaipwvc-01}. 

\begin{theorem}
  \textsc{MinSPED} can be 2-approximated in time quadratic in the
  number of crossings of the given geometric graph.
\end{theorem}

\begin{pf}
  Given an instance $G$ of \textsc{MinSPED}, we construct an
  instance~$\varphi$ of \textsc{MinW2Sat} as follows.
Let~$e$ be an edge of~$G$ with $k$ crossings.  Then~$e$ is split into $k+1$ pairs of edge segments $e_0, \dots, e_k$ as shown in \figurename~\ref{fig:approx}.
If we order the edges that cross $e$ in increasing order of the
distance of their crossing point to the closer endpoint of $e$, we can
assign each segment pair $e_i$ for $i \ge 1$ to the $i$th edge $f^i$
crossing~$e$, in this order.
We also say that edge $f^i$ \emph{induces} segment pair $e_i$.
Any valid maximal (non-extensible) partial edge drawing of $e$ is the union $\bigcup_{i=0}^j e_i$ of all pairs of edge segments up to some index $j\le k$.

\begin{figure}[tb]
  \centering
  \includegraphics{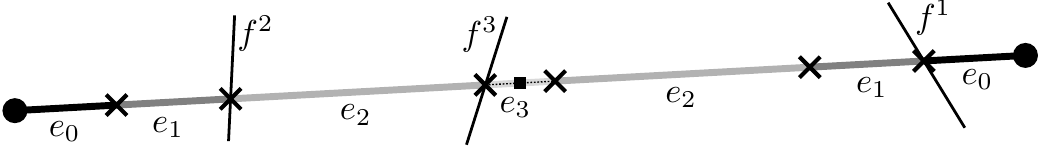}
  \caption{Edge $e$ is split into four pairs of edge segments; 
    pairs are labeled equally.} 
  \label{fig:approx}
\end{figure}

We model all pairs of (induced) edge segments as truth variables $\hat{e}_1,
\dots, \hat{e}_k$ with the interpretation that the pair~$e_i$ is
\emph{not} drawn if $\hat{e}_i = \mathit{true}$.  
The pair~$e_0$ is always drawn.
For $i=1, \dots, k$, we introduce the clause $(\neg \hat{e}_{i+1}
\Rightarrow \neg \hat{e}_i) \equiv (\hat{e}_{i+1} \lor \neg \hat{e}_i)$.   
This models that 
$e_{i+1}$ can only be drawn if~$e_i$ is drawn.
Moreover, for every crossing between two edges~$e$ and~$f$,
we introduce the clause $(e_i \lor f_j)$, where $e_i$ is the segment pair
of~$e$ induced by~$f$ and $f_j$ is the segment pair of~$f$ induced
by~$e$. 
This simply means that at least one of the two induced segment pairs
is not drawn and thus the crossing is avoided.

Now we assign a weight~$w_{e,i}$ to each variable~$\hat{e}_i$, which
is either the absolute length~$|e_i|$ of~$e_i$ if we are interested in
ink, or the relative length $|e_i|/(2|e|)$ if we are interested in
relative stub lengths ($\delta$).
Then minimizing the value $\sum_{\hat{e}_i \in \text{Var}(\varphi)} w_{e,i}
\hat{e}_i$ over all valid variable assignments minimizes the weight of
the erased parts of the edges in the given geometric graph

The 2-approximation algorithm for \textsc{MinW2Sat} yields a
2-approximation for the problem to erase the minimum ink from the
given straight-line drawing of~$G$.  It runs in $O(vc)=O(I^2)
\subseteq O(m^4)$ time since our 2-SAT formula has $O(I) \subseteq
O(m^2)$ variables and clauses, where $m$ is the number of edges of~$G$
and~$I$ is the number of intersections in the drawing of~$G$.
\end{pf}

If we encode the primal problem (maximize ink) using 2SAT, we cannot
hope for a similar positive result.  The reason is that the tool that
we would need, namely an algorithm for the problem \textsc{MaxW2Sat}
dual to \textsc{MinW2Sat} would also solve maximum independent set
(\textsc{MIS}).  For MIS, however, no $(n^{1-\eps})$-approximation
exists unless ${\cal NP}={\cal ZPP}$ \cite{h-chan1e-AM99}.

To see that \textsc{MaxW2Sat} can be used to encode MIS, use a
variable~$\hat v$ for each vertex~$v$ of the given (graph)
instance~$G$ of MIS and, for each edge $\{u,v\}$ of~$G$, the clause
$(\hat{u} \vee \hat{v})$.  Let $\varphi$ be the conjunction of all
such clauses.  Then finding a satisfying truth assignment
for~$\varphi$ that maximizes the number of false variables (i.e.,
all variable weights are~1) is equivalent to finding a maximum
independent set in~$G$.
Note that this does \emph{not} mean that maximizing ink is as hard
to approximate as MIS.

\bigskip
\noindent\textbf{Acknowledgments.}  We thank Ferran Hurtado and Yoshio
Okamoto for invaluable pointers to results in discrete geometry.  We
thank Emilio Di Giacomo, Antonios Symvonis, Henk Meijer, Ulrik
Brandes, and Ga{\v s}per Fijav{\v z} for helpful hints and intense
discussions. 
We thank Jarek Byrka for the link between ink maximization and MIS.
Thanks to Thomas van Dijk for drawing Figure~\ref{fig:example} (and
implementing the ILP behind it).

\bibliographystyle{alpha}
\bibliography{abbrv,lncs,edgebundling,ped}

\end{document}